\documentclass[runningheads]{CMSIM}

\usepackage{graphicx} 

\usepackage{amsmath,amssymb}

\renewcommand{\thefootnote}

\title*{Fisher Information Perspective of \\ Pauli's Electron}

\titlerunning{\it Fisher Information Perspective of Pauli's Electron}

\author{
Asher Yahalom
}

\authorrunning{\it Asher Yahalom}

\institute{
Ariel University, Ariel 40700, Israel\\
(E-mail: {\tt asya@ariel.ac.il})
}

\begin{document}
\thispagestyle{empty}
\maketitle
\setlength{\leftskip}{0pt}
\setlength{\headsep}{16pt}
\footnote{\begin{tabular}{p{11.2cm}r}
\small {\it $15^{th}$CHAOS Conference Proceedings, 14 - 17 June 2022, Athens, Greece} \\
\small C. H. Skiadas (Ed)\\
\small \textcopyright {} 2022 ISAST & \includegraphics[scale=0.35]{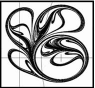}
 \end{tabular}}
\begin{abstract}
A non relativistic electron with a spin is described by Pauli's equation.
It has been shown that this system can be interpreted as a vortical fluid which has both
similarities and differences with classical ideal flows. Moreover, it was demonstrated that the internal energy of the spin fluid can partially be interpreted in terms of Fisher Information. While previous work on the subject has mainly ignored electromagnetic fields which are represented by a vector potential here we remove this limitation and study the system under general electromagnetic interaction.
\end{abstract}

\newcommand{\beq} {\begin{equation}}
\newcommand{\enq} {\end{equation}}
\newcommand{\ber} {\begin {eqnarray}}
\newcommand{\enr} {\end {eqnarray}}
\newcommand{\eq} {equation}
\newcommand{\eqn} {equation }
\newcommand{\eqs} {equations }
\newcommand{\ens} {equations}
\newcommand {\er}[1] {equation (\ref{#1}) }
\newcommand {\ern}[1] {equation (\ref{#1})}
\newcommand {\ers}[1] {equations (\ref{#1})}
\newcommand {\Er}[1] {Equation (\ref{#1}) }
\newcommand{\br} {\bar{r}}
\newcommand{\tnu} {\tilde{\nu}}
\newcommand{\rhm}  {{\rho \mu}}
\newcommand{\sr}  {{\sigma \rho}}
\newcommand{\bh}  {{\bar h}}
\newcommand {\Sc} {Schr\"{o}dinger}
\newcommand {\SE} {Schr\"{o}dinger equation }
\newcommand {\bR} {{\bf R}}
\newcommand {\bX} {{\bf X}}
\newcommand{\ce}  {continuity equation }
\newcommand{\ces} {continuity equations }
\newcommand{\hje} {Hamilton-Jacobi equation }
\newcommand{\hjes} {Hamilton-Jacobi equations }
\newcommand{\bp}  {\bar{\psi}}
\newcommand{\va}  {\vec \alpha}

\section {Introduction}

The Copenhagen interpretation of quantum mechanics is probably the most prevalent approach. This approach  defies any ontology to quantum theory and declares it to be completely epistemological in accordance to
the Kantian \cite{Kant} conception of reality. However, in addition to this approach we see the development of another school that believed in the realism of the wave function. This approach that was championed by Einstein and Bohm \cite{Bohm,Holland,DuTe} led to other interpretations of quantum mechanics among them the fluid interpretation due to do Madelung \cite{Madelung,Complex} which interpreted the modulus square as the fluid density and the phase as a potential of a velocity field. However, this model was limited to spin less electrons and could not take into account a complete set of  electron attributes even for non relativistic electrons.

A spin dependent non relativistic quantum equation was first introduced by Wolfgang Pauli in 1927 \cite{Pauli}. This equation
contained a Hamiltonian which is a two dimensional operator matrix. Such two dimensional operator Hamiltonians were later found useful
for many systems that required quantum modelling among them molecules and solids. Such two dimensional operator matrix Hamiltonians are abundant in the literature (\cite{EYB1} - \cite{EY8}). The question now arises wether such a theory admits a fluid dynamical interpretation. This question seems of paramount importance as the proponents of the non-realistic Copenhagen interpretation of quantum mechanics usually use the concept of spin as a proof that some elements of nature are inherently quantum and have no classical analogue or interpretation.
A Bohmian interpretation of the Pauli equation was given by Holland and others \cite{Holland}, however, the relation of this equation
to fluid dynamics and the concept of spin vorticity were not introduced. This situation was amended in a recent paper describing spin fluid dynamics \cite{Spflu}.

The formulation of Pauli's theory in terms of a fluid theory leads us directly to the nineteenth century work of Clebsch \cite{Clebsch1,Clebsch2} and the variational formulation of fluid dynamics. Variational principles for barotropic fluid dynamics are well known.
A four function variational formulation of Eulerian barotropic fluid
dynamics  was derived by  Clebsch \cite{Clebsch1,Clebsch2} and later by Davidov \cite{Davidov}
who's main motivation was to quantize fluid dynamics. Since the work was written in Russian, it was unknown in the west.
Lagrangian fluid  dynamics (as opposed to Eulerian fluid
 dynamics) was formulated through a variational principle by Eckart \cite{Eckart}.
Initial western attempts to formulate Eulerian fluid dynamics in terms of a variational
principle, were described by Herivel \cite{Herivel}, Serrin
\cite{Serrin} and Lin \cite{Lin}. However, the variational principles
developed by the above authors were very cumbersome containing
quite a few "Lagrange multipliers" and "potentials". The range of
the total number of independent functions in the above
formulations ranges from eleven to seven which exceeds by many the
four functions appearing in the Eulerian and continuity equations
of a barotropic flow. And therefore did not have any practical use
or applications. Seliger \& Whitham \cite{Seliger} have developed
a variational formalism depending on only four
variables for barotropic flow and thus repeated the work of Davidov's \cite{Davidov} which
they were unaware of. Lynden-Bell \& Katz \cite{LynanKatz} have described a variational principle in
terms of two functions the load $\lambda$ and density $\rho$.
However, their formalism contains an implicit definition for the velocity $\vec v$
such that one is required to solve a partial differential equation in order
to obtain both $\vec v$ in terms of $\rho$ and $\lambda$ as well as its variations.
Much the same criticism holds for their general variational for non-barotropic flows \cite{KatzLyndeb}.
Yahalom \& Lynden-Bell \cite{YahLyndeb} overcame this limitation by paying the price of adding an additional
single function. Their formalism allowed arbitrary variations and the definition of $\vec v$ is explicit.

A fundamental problem in the fluid mechanical interpretation of quantum mechanics still exist. This refers to the meaning of thermodynamic quantities which are part of fluid mechanics. In thermodynamics Concepts like specific enthalpy,
 pressure and temperature are
 derivatives of the specific internal energy which is given in terms of the equation of state as function of entropy and density.
 The internal energy is a part of any Lagrangian density attempting to describe fluid dynamics.
 The form of the internal energy can in principle be explained on the basis of the microscopic composition of the fluid, that is the atoms and
 molecules from which the fluid is composed and their interactions using statistical mechanics. However, the quantum fluid has no microscopic structure
 and yet analysis of the equations of both the spin less \cite{Madelung,Complex} and spin \cite{Spflu} quantum fluid dynamics shows that terms analogue
 to internal energies appear in both cases. The question then arises where do those internal energies come from, surely one would not suggest that the
 quantum fluid has a microscopic sub structure as this will defy the conception of the electron as a fundamental particle. The answer to this question
 seems to come from an entirely different discipline of measurement theory \cite{Fisher,Fisherspin}. Fisher information a basic notion of measurement theory
 is a measure of the quality of
 the  measurement of any quantity. It was shown \cite{Fisherspin} that this concept is proportional to the internal energy of a spin less electron and can explain parts of the internal energy of an electron with spin. An attempt to unify most physical theories using Fisher information is described in a book by Frieden \cite{Frieden}. It was conjectured
 \cite{Fisherspin2} that there exist a velocity field such that the Fisher information will given a complete explanation for the spin fluid internal energy. It was also suggested that one may define comoving scalar fields as in ideal fluid mechanics, however, this was only demonstrated implicitly but not explicitly. A common feature of previous work on the fluid \& Fisher information interpretation of quantum mechanics, is the negligence of electromagnetic interaction thus setting the vector potential to zero. This makes sense as the classical ideal fluids discussed in the literature are not charged. Hence, in order to make the comparison easier to comprehend the vector potential should be neglected. However, one cannot claim a complete description of quantum mechanics lacking a vector potential thus ignoring important quantum phenomena such as the Zeeman effect which depends on a vector potential through the magnetic field.

 We will begin this paper by introducing a variational principle for a charged classical particle with a vector potential interaction and a system of the same. This will be followed by the Eckart \cite{Eckart} Lagrangian variational principles generalized for a charged fluid. We then introduce an Eulerian-Clesch variational principle for a charged fluid. Next we will introduce the Fisher information and the concept of probability amplitude.   This will be followed by a discussion of
 \SE with a non trivial vector potential and its interpretation in terms of Madelung fluid dynamics. The variational principle of a charged Madelung fluid dynamics
 will be described and its relations to Fisher information will be elucidated.  Then we introduce Pauli's equation
 with a vector potential and interpret it in terms of generalized Clebsch variables and discuss the similarities and differences between a charge Clebsch flow  and spin fluid dynamics and their variational principles. Finally the concept of Fisher information will be introduced  in the framework of spin fluid dynamics.

 \section{A Classical Charged Particle}

 Consider a classical particle with the coordinates $\vec x (t)$, mass $m$ and charge $e$ interacting with a given electromagnetic vector potential $\vec A (\vec x,t)$ and scalar potential $\varphi (\vec x,t)$.
 We will not be interested in the effects of the particle on the field and thus consider the field as "external". The action of the said particle is:
 \ber
 {\cal A} &=& \int_{t1}^{t2} L dt, \qquad L = L_0 + L_i
 \nonumber \\
 L_0 &\equiv& \frac{1}{2} m v^2, \qquad L_i \equiv e(\vec A \cdot \vec v - \varphi), \qquad
  \vec v \equiv \frac{d \vec x}{dt} \equiv \dot{\vec x}, \quad v =|\vec v|.
 \label{classparticleact}
 \enr
The variation of the two parts of the Lagrangian are given by:
\beq
 \delta L_0 =  m \dot{\vec x  } \cdot \delta \dot{\vec x} =
 \frac{d (m \vec v \cdot \delta \vec x)}{dt} - m \dot {\vec v} \cdot \delta \vec x
 \label{delLf}
 \enq
\ber
 \delta L_i &=&  e \left( \delta \vec A \cdot \vec v + \vec A \cdot \delta \dot{\vec x} -
 \delta \varphi \right)
 \nonumber \\
 &=& e \left( \partial_k \vec A \cdot \vec v \delta x_k +
  \frac{d (\vec A \cdot \delta \vec x)}{dt} - \delta \vec x \cdot \frac{d \vec A}{dt}
  - \vec \nabla \varphi \cdot \delta \vec x\right),
 \label{delLi}
 \enr
in the above $\partial_k \equiv \frac{\partial }{\partial x_k}$ and
$\vec \nabla \equiv (\frac{\partial }{\partial x},\frac{\partial }{\partial y},\frac{\partial }{\partial z}) \equiv (\frac{\partial }{\partial x_1},\frac{\partial }{\partial x_2},\frac{\partial }{\partial x_3})$.
We use the Einstein summation convention in which a Latin index (say $k,l$) takes one of the values $k,l \in [1,2,3]$. We may write the total time derivative of $\vec A$ as:
\beq
\frac{d \vec A (\vec x (t),t)}{dt} =  \partial_t \vec A + v_l \partial_l \vec A,
\qquad  \partial_t \equiv \frac{\partial }{\partial t}.
\label{Ader}
\enq
Thus the variation $\delta L_i$ can be written in the following form:
\beq
 \delta L_i
 =\frac{d (e \vec A \cdot \delta \vec x)}{dt} +
  e \left[ \left(\partial_k  A_l -\partial_l  A_k \right) v_l -  \partial_t A_k
    - \partial_k \varphi \right] \delta x_k,
 \label{delLi2}
 \enq
Defining the electric and magnetic fields in the standard way:
\beq
\vec B \equiv \vec \nabla \times \vec A, \qquad \vec E \equiv  -\partial_t \vec A - \vec \nabla \varphi,
\label{EB}
 \enq
it follows that:
 \beq
\epsilon_{kln} B_n = \partial_k  A_l -\partial_l  A_k  , \qquad  E_k =  -\partial_t  A_k - \partial_k \varphi,
\label{EB2}
 \enq
in which $ \epsilon_{kln}$ is the three index antisymmetric tensor. Thus we may write $\delta L_i$ as:
 \beq
 \delta L_i
 =\frac{d (e \vec A \cdot \delta \vec x)}{dt} +
  e \left[ \epsilon_{kln} v_l B_n + E_k \right] \delta x_k
  = \frac{d (e \vec A \cdot \delta \vec x)}{dt} +
  e \left[ \vec v \times \vec B + \vec E\right] \cdot \delta \vec x.
 \label{delLi3}
 \enq
 We use the standard definition of the Lorentz force (MKS units):
 \beq
 \vec F_L \equiv e \left[ \vec v \times \vec B + \vec E\right]
 \label{Lor}
 \enq
 to write:
 \beq
 \delta L_i   = \frac{d (e \vec A \cdot \delta \vec x)}{dt} + \vec F_L \cdot \delta \vec x.
 \label{delLi4}
 \enq
 Combining the variation of $L_i$ given in \ern{delLi4} and the variation of $L_0$ given in \ern{delLf}, it follows from \ern{classparticleact} that the variation of $L$ is:
  \beq
  \delta L =\delta L_0 + \delta L_i =
   \frac{d \left( (m \vec v + e \vec A) \cdot \delta \vec x\right)}{dt} +(-m \dot{\vec v} +\vec F_L) \cdot \delta \vec x.
 \label{delL1}
 \enq
 Thus the variation of the action is:
  \beq
  \delta A =\int_{t1}^{t2} \delta L dt =
  \left. (m \vec v + e \vec A) \cdot \delta \vec x \right|_{t1}^{t2}
   - \int_{t1}^{t2}(m \dot{\vec v} - \vec F_L) \cdot \delta \vec x dt.
 \label{delA1}
 \enq
 Since the classical trajectory is such that the variation of the action on it vanishes for
 a small modification of the trajectory $\delta \vec x$ that vanishes at $t1$ and $t2$ but is otherwise arbitrary it follows that:
 \beq
  m \dot{\vec v} = \vec F_L = e \left[\vec v \times \vec B + \vec E \right]
  \Rightarrow \dot{\vec v} = \frac{e}{m} \left[\vec v \times \vec B + \vec E \right].
 \label{equamotion}
 \enq
 Thus the dynamics of a classical particle in a given electric and magnetic field is described by a single number, the ratio between its charge and mass:
  \beq
  k \equiv \frac{e}{m} \qquad \Rightarrow \qquad \dot{\vec v} = k \left[\vec v \times \vec B + \vec E \right].
 \label{equamotion2}
 \enq
The reader is reminded that the connection between the electromagnetic potentials and the fields is not unique. Indeed performing  a gauge transformation to obtain a new set of potentials:
\beq
  \vec A' = \vec A + \vec \nabla \Lambda, \qquad \varphi' = \varphi - \partial_t \Lambda.
 \label{gauge}
 \enq
we obtain the same fields:
\beq
 \vec B' = \vec \nabla \times \vec A' = \vec \nabla \times \vec A = \vec B, \qquad
 \vec E' =   -\partial_t \vec A' - \vec \nabla \varphi' =  -\partial_t \vec A - \vec \nabla \varphi = \vec E.
 \label{gauge2}
 \enq
 For a system of $N$ particles each with an index $j \in [1-N]$, a corresponding mass $m_j$, charge  $e_j$, position vector $\vec x_j$ and velocity $\vec v_j \equiv \frac{d \vec x_j}{dt}$ the action and Lagrangian for each point particle are as follows:
 \ber
 {\cal A}_j &=& \int_{t1}^{t2} L_j dt, \qquad L_j = L_{0j} + L_{ij}
 \nonumber \\
 L_{0j} &\equiv& \frac{1}{2} m_j v_j^2, \qquad L_{ij} \equiv e_j(\vec A (\vec x_j,t) \cdot \vec v_j - \varphi (\vec x_j,t)).
 \label{classparticlej}
 \enr
 The action and Lagrangian of the system of particles is:
\beq
 {\cal A}_s = \int_{t1}^{t2} L_s dt, \qquad L_s = \sum_{j=1}^{N} L_j.
  \label{classparticlesys}
 \enq
 The variational analysis follows the same lines as for a single particle and we obtain a set of equations of the form:
 \beq
 \dot{\vec v}_j = \frac{e_j}{m_j} \left[\vec v_j \times \vec B (\vec x_j,t) + \vec E(\vec x_j,t) \right], \qquad j \in [1-N].
 \label{equamotionj}
 \enq

\section{A Classical Charged Fluid - the Lagrangian Approach}

The dynamics of the fluid is determined by its composition and the forces acting on it. The fluid is made of "fluid elements" \cite{Eckart,Bertherton}, practically a "fluid element" is a point particle which has an infinitesimal mass $d M_{\vec \alpha}$, infinitesimal charge $d Q_{\vec \alpha}$,  position vector $\vec x (\vec \alpha,t)$ and velocity $\vec v (\vec \alpha,t) \equiv \frac{d \vec x (\vec \alpha,t)}{dt}$. Here the continuous vector label $\vec \alpha$ replaces the discrete index $j$ of the previous section. As the "fluid element" is not truly a point particle it has also an infinitesimal volume $d V_{\vec \alpha}$, infinitesimal entropy $d S_{\vec \alpha}$, and an infinitesimal internal energy $d E_{in~\vec \alpha}$. The action and Lagrangian for each "fluid element" are
according to \ern{classparticleact} as follows:
\ber
 d {\cal A}_{\va} &=& \int_{t1}^{t2} d L_{\va} dt, \qquad d L_{\va} = d L_{0\va} + d L_{i\va}
 \nonumber \\
 d L_{0\va} &\equiv& \frac{1}{2} d M_{\va}~ v(\va,t)^2 - d E_{in~\vec \alpha} ,
 \nonumber \\
  d L_{i\va} &\equiv& d Q_{\va} \left(\vec A (\vec x (\va,t),t) \cdot \vec v(\va,t) - \varphi(\vec x (\va,t),t)\right).
 \label{classparticleactcf}
 \enr
all the above quantities are calculated for a specific value of the label $\vec \alpha$, while the action and Lagrangian of the entire fluid, should be summed (or integrated) over all possible
$\vec \alpha$'s. That is:
\ber
L &=& \int_{\va}  d L_{\vec \alpha}
\nonumber \\
{\cal A} &=& \int_{\va} d {\cal A}_{\vec \alpha} = \int_{t1}^{t2} \int_{\va}  d L_{\vec \alpha} dt
 =  \int_{t1}^{t2} L dt.
 \label{fluac1}
 \enr
 It is customary to define densities for the Lagrangian, mass and charge of every fluid element as
 follows:
 \beq
 {\cal L}_{\va} \equiv \frac{d L_{\va}}{d V_{\va}}, \quad
  \rho_{\va} \equiv \frac{d M_{\va}}{d V_{\va}} , \quad
\rho_{c\va} \equiv \frac{d Q_{\va}}{d V_{\va}}
\label{dens}
\enq
Each of the above quantities may be thought of as a function of the location $\vec x$, where the "fluid element" labelled $\va$ happens to be in time $t$, for example:
\beq
 \rho (\vec x, t) \equiv \rho (\vec x (\va,t), t) \equiv \rho_{\va} (t)
\label{dens2}
\enq
It is also customary to define the specific internal energy $\varepsilon_{\va}$ as follows:
\beq
 \varepsilon_{\va} \equiv \frac{d E_{in~\va}}{d M_{\va}} \quad \Rightarrow  \quad
 \rho_{\va} \varepsilon_{\va} = \frac{d M_{\va}}{d V_{\va}} \frac{d E_{in~\va}}{d M_{\va}} =
 \frac{d E_{in~\va}}{d V_{\va}}
\label{specifint}
\enq
Thus we can write the following equations for the Lagrangian density:
\ber
 {\cal L}_{\va} &=& \frac{d L_{\va}}{d V_{\va}}=
   \frac{{d L}_{0\va}} {d V_{\va}} + \frac{{d L}_{i\va}} {d V_{\va}}
 = {\cal L}_{0\va} +  {\cal L}_{i\va}
 \nonumber \\
 {\cal L}_{0\va} &\equiv& \frac{1}{2} \rho_{\va} v (\va,t)^2 - \rho_{\va} \varepsilon_{\va} ,
  \nonumber \\
  {\cal L}_{i\va} &\equiv& \rho_{c\va} \left(\vec A (\vec x (\va,t),t) \cdot \vec v (\va,t) - \varphi (\vec x (\va,t),t)\right).
 \label{lagdensity}
 \enr
 The above expression allows us to write the Lagrangian as a spatial integral:
 \beq
L = \int_{\va}  d L_{\vec \alpha} = \int_{\va} {\cal L}_{\va} d V_{\va}
= \int {\cal L} (\vec x,t) d^3 x
\label{Lspatint}
\enq
which will be important for later sections of the current paper. Returning now to the variational analysis we introduce the symbols $\Delta \vec x_{\va} \equiv \vec \xi_{\va}$ to indicate a variation of the trajectory $ \vec x_{\va} (t)$ (we reserve the symbol $\delta$  in the fluid context, to a different kind of variation, the Eulerian variation to be described in the next section). In an ideal fluid the "fluid element" does exchange mass, nor electric charge, nor heat with other fluid elements, so it follows that:
 \beq
\Delta d M_{\vec \alpha} = \Delta d Q_{\vec \alpha} = \Delta d S_{\vec \alpha}  =0.
\label{consvl}
\enq
Moreover, according to thermodynamics a change in the internal energy of a "fluid element" satisfies
the equation:
\beq
\Delta d E_{in~\vec \alpha} = T_{\va} \Delta dS_{\vec \alpha} - P_{\va} \Delta dV_{\vec \alpha},
\label{thermo}
\enq
the first term describes the heating energy gained by the "fluid element" while the second terms describes the work done by the "fluid element" on neighbouring elements. $T_{\va}$ is the temperature of the "fluid element" and $P_{\va}$ is the pressure of the same. As the mass of the fluid element does not change we may divide the above expression by $d M_{\vec \alpha}$ to obtain
the variation of the specific energy as follows:
\ber
\Delta \varepsilon_{\va}  &=& \Delta\frac{d E_{in~\vec \alpha}}{d M_{\va}}
 =  T_{\va} \Delta\frac{ dS_{\vec \alpha}}{d M_{\va}} - P_{\va} \Delta \frac{dV_{\vec \alpha}}{d M_{\va}}
 \nonumber \\
 &=& T_{\va} \Delta s_{\va} - P_{\va} \Delta \frac{1}{\rho_{\va}}
 = T_{\va} \Delta s_{\va} + \frac{P_{\va}}{\rho_{\va}^2} \Delta \rho_{\va}. \qquad
 s_{\va} \equiv \frac{ dS_{\vec \alpha}}{d M_{\va}}
\label{thermo2b}
\enr
in which $s_{\va}$ is the specific entropy of the fluid element. It follows that:
\beq
\frac{\partial \varepsilon }{\partial s} = T, \qquad
\frac{\partial \varepsilon }{\partial \rho} = \frac{P}{\rho^2}.
\label{thermo2c}
\enq
Another important thermodynamic quantity that we will use later is the Enthalpy defined for
a fluid element as:
\beq
 dW_{\va} = d E_{in~\vec \alpha} + P_{\va} d V_{\va}.
\label{thermo2d}
\enq
and the specific enthalpy:
\beq
 w_{\va} =\frac{dW_{\va}}{d M_{\va}}= \frac{dE_{in~\vec \alpha}}{d M_{\va}}
 +P_{\va} \frac{dV_{\va}}{d M_{\va}} = \varepsilon_{\va} + \frac{P_{\va}}{\rho_{\va}}.
\label{thermo2e}
\enq
Combining the above result with \ern{thermo2c} it follows that:
\beq
 w = \varepsilon + \frac{P}{\rho} = \varepsilon + \rho \frac{\partial \varepsilon }{\partial \rho}
 = \frac{\partial (\rho \varepsilon) }{\partial \rho}.
\label{thermo2f}
\enq
Moreover:
\beq
 \frac{\partial w }{\partial \rho} = \frac{\partial (\varepsilon + \frac{P}{\rho})}{\partial \rho}
  = - \frac{P}{\rho^2} + \frac{1}{\rho} \frac{\partial P }{\partial \rho}+ \frac{\partial \varepsilon }{\partial \rho} =
  - \frac{P}{\rho^2} + \frac{1}{\rho} \frac{\partial P }{\partial \rho}+ \frac{P}{\rho^2} =
   \frac{1}{\rho} \frac{\partial P }{\partial \rho}.
\label{thermo2g}
\enq
As we assume an ideal fluid, there is no heat conduction or heat radiation, and thus heat can only be moved around along the trajectory of the "fluid elements", that is only convection is taken into account. Thus $\Delta d S_{\vec \alpha}  =0$
and we have:
\beq
\Delta d E_{in~\vec \alpha} = - P \Delta dV_{\vec \alpha}.
\label{thermo2}
\enq
Our next step would to be to evaluate the variation of the volume element. Suppose a time $t$ the
volume of the fluid element labelled by $\va$ is described as:
\beq
 dV_{\va,t} = d^3 x(\va, t)
\label{volelem}
\enq
Using the Jacobian determinant we may relate this to the same element at $t=0$:
\beq
  d^3 x(\va, t)  = J  d^3 x(\va, 0), \qquad
  J \equiv \vec \nabla_0 x_1 \cdot (\vec \nabla_0 x_2 \times \vec \nabla_0 x_3)
\label{volelem2}
\enq
In which $\vec \nabla_0$ is taken with respect to the coordinates of the fluid elements at $t=0$: $\vec \nabla_0 \equiv (\frac{\partial }{\partial x(\va,0)_1},\frac{\partial }{\partial x(\va,0)_2},\frac{\partial }{\partial x(\va,0)_3})$. As both the actual and varied "fluid element" trajectories start at the same point it follows that:
\ber
\Delta dV_{\va,t} &=& \Delta d^3 x(\va, t)  = \Delta J ~ d^3 x(\va, 0)  = \frac{\Delta J}{J} d^3 x(\va, t) = \frac{\Delta J}{J}  dV_{\va,t},
\nonumber \\
(\Delta d^3 x(\va, 0) &=& 0).
\label{volelem3}
\enr
The variation of $J$ can be easily calculated as:
\beq
    \Delta J = \vec \nabla_0 \Delta x_1 \cdot (\vec \nabla_0 x_2 \times \vec \nabla_0 x_3)
    + \vec \nabla_0  x_1 \cdot (\vec \nabla_0 \Delta x_2 \times \vec \nabla_0 x_3)
    + \vec \nabla_0  x_1 \cdot (\vec \nabla_0  x_2 \times \vec \nabla_0 \Delta x_3),
\label{volelem4}
\enq
Now:
\ber
& & \vec \nabla_0 \Delta x_1 \cdot (\vec \nabla_0 x_2 \times \vec \nabla_0 x_3)
= \vec \nabla_0 \xi_1 \cdot (\vec \nabla_0 x_2 \times \vec \nabla_0 x_3)
\nonumber \\
&=& \partial_k \xi_1 \vec \nabla_0 x_k \cdot (\vec \nabla_0 x_2 \times \vec \nabla_0 x_3)
= \partial_1 \xi_1 \vec \nabla_0 x_1 \cdot (\vec \nabla_0 x_2 \times \vec \nabla_0 x_3)=
\partial_1 \xi_1 J.
\nonumber \\
& & \vec \nabla_0  x_1 \cdot (\vec \nabla_0 \Delta x_2 \times \vec \nabla_0 x_3)
= \vec \nabla_0 x_1 \cdot (\vec \nabla_0 \xi_2 \times \vec \nabla_0 x_3)
\nonumber \\
&=& \partial_k \xi_2 \vec \nabla_0 x_1 \cdot (\vec \nabla_0 x_k \times \vec \nabla_0 x_3)
= \partial_2 \xi_2 \vec \nabla_0 x_1 \cdot (\vec \nabla_0 x_2 \times \vec \nabla_0 x_3)=
\partial_2 \xi_2 J.
\nonumber \\
& & \vec \nabla_0  x_1 \cdot (\vec \nabla_0  x_2 \times \vec \nabla_0 \Delta x_3)
= \vec \nabla_0 x_1 \cdot (\vec \nabla_0 x_2 \times \vec \nabla_0 \xi_3)
\nonumber \\
&=& \partial_k \xi_3 \vec \nabla_0 x_1 \cdot (\vec \nabla_0 x_2 \times \vec \nabla_0 x_k)
= \partial_3 \xi_3 \vec \nabla_0 x_1 \cdot (\vec \nabla_0 x_2 \times \vec \nabla_0 x_3)=
\partial_3 \xi_3 J.
\label{volelem5}
\enr
Combining the above results, it follows that:
\beq
    \Delta J = \partial_1 \xi_1 J + \partial_2 \xi_2 J + \partial_3 \xi_3 J =
     \vec \nabla \cdot \vec \xi~ J.
\label{volelem6}
\enq
Which allows us to calculate the variation of the volume of the "fluid element":
\beq
\Delta dV_{\va,t}  = \vec \nabla \cdot \vec \xi ~ dV_{\va,t}.
\label{volelem7}
\enq
And thus the variation of the internal energy is:
\beq
\Delta d E_{in~\vec \alpha} = - P \vec \nabla \cdot \vec \xi ~ dV_{\va,t}.
\label{thermo3}
\enq
The internal energy is the only novel element with respect to the single particle scenario and system of particles scenario described in the previous section, thus the rest of the variation analysis is straight forward. Varying \ern{classparticleactcf} we obtain:
\ber
 \Delta d {\cal A}_{\va} &=& \int_{t1}^{t2} \Delta d L_{\va} dt, \qquad \Delta d L_{\va} = \Delta d L_{0\va} + \Delta d L_{i\va}
 \nonumber \\
\Delta d L_{0\va} &=&  d M_{\va}~\vec v(\va,t) \cdot \Delta \vec v(\va,t)   - \Delta d E_{in~\vec \alpha} ,
 \nonumber \\
  \Delta d L_{i\va} &=& d Q_{\va} \left(\Delta \vec A (\vec x (\va,t),t) \cdot \vec v(\va,t)
  + \vec A (\vec x (\va,t),t) \cdot \Delta \vec v(\va,t) \right.
  \nonumber \\
  &-& \left. \Delta \varphi(\vec x (\va),t)\right).
 \label{varclassparticleactcf}
 \enr
Notice that:
\beq
\Delta \vec v(\va,t) =  \Delta \frac{d \vec x (\vec \alpha,t)}{dt} = \frac{d \Delta \vec x (\vec \alpha,t)}{dt} = \frac{d \vec \xi (\vec \alpha,t)}{dt}.
\label{vvar}
\enq
After some steps which are described prior to \ern{delLf} we obtain the variation of $d L_{0\va}$:
\beq
\Delta d L_{0\va} =
 \frac{d (d M_{\va} \vec v_{\va} \cdot \vec \xi_{\va})}{dt} - d M_{\va} \frac{d {\vec v}_{\va} }{dt} \cdot \vec \xi_{\va} + P \vec \nabla \cdot \vec \xi_{\va} ~ dV_{\va,t}.
 \label{delLffl}
 \enq
Similarly the analogue equations of \ern{Lor} and \ern{delLi4} are:
\beq
 d \vec F_{L\va} \equiv d Q_{\va} \left[ \vec v_{\va} \times \vec B (\vec x (\va,t),t) + \vec E (\vec x (\va,t),t) \right]
 \label{Lorfl}
 \enq
and:
 \beq
 \Delta d L_{i\va} = \frac{d (d Q_{\va} \vec A (\vec x (\va,t),t) \cdot \vec \xi_{\va})}{dt} +
  d \vec F_{L\va} \cdot \vec \xi_{\va}.
 \label{delLi4fl}
 \enq
 The variation of the action of a single fluid element is thus:
 \ber
  \Delta d {\cal A}_{\va} &=& \int_{t1}^{t2} \Delta dL_{\va} dt =
  \left. (d M_{\va} \vec v (\va,t) + d Q_{\va} \vec A (\vec x (\va,t),t)) \cdot \vec \xi_{\va} \right|_{t1}^{t2}
  \nonumber \\
   &-& \int_{t1}^{t2}(d M_{\va} \frac{d \vec v (\va,t)}{dt}\cdot \vec \xi_{\va} - d \vec F_{L\va} \cdot \vec \xi_{\va} - P \vec \nabla \cdot \vec \xi_{\va} ~ dV_{\va,t} )  dt.
 \label{delA1fl}
 \enr
The variation of the total action of the fluid is thus:
\ber
\Delta {\cal A} &=& \int_{\va} d {\cal A}_{\vec \alpha} =
\left. \int_{\va} (d M_{\va} \vec v (\va,t) + d Q_{\va} \vec A (\vec x (\va,t),t)) \cdot \vec \xi_{\va} \right|_{t1}^{t2}
  \nonumber \\
   &-& \int_{t1}^{t2}\int_{\va} (d M_{\va} \frac{d \vec v (\va,t)}{dt}\cdot \vec \xi_{\va} - d \vec F_{L\va} \cdot \vec \xi_{\va} - P \vec \nabla \cdot \vec \xi_{\va} ~ dV_{\va} )  dt.
 \label{varAfl2}
 \enr
 Now according to \ern{dens} we may write:
 \beq
 d M_{\va} = \rho_{\va}~dV_{\va}, \qquad d Q_{\va} = \rho_{c\va}~dV_{\va}
 \label{dvq}
 \enq
using the above relations we may turn the $\va$ integral into a volume integral and thus write
the variation of the fluid action in which we suppress the $\va$ labels:
\beq
\Delta {\cal A} =
\left. \int (\rho \vec v  + \rho_c \vec A)  \cdot \vec \xi dV \right|_{t1}^{t2}
  - \int_{t1}^{t2}\int (\rho \frac{d \vec v}{dt}\cdot \vec \xi - \vec f_{L} \cdot \vec \xi - P \vec \nabla \cdot \vec \xi)  dV  dt.
 \label{varAfl3}
 \enq
in the above we introduced the Lorentz force density:
\beq
\vec f_{L\va} \equiv \frac{d \vec F_{L\va}}{dV_{\va}} = \rho_{c\va} \left[ \vec v_{\va} \times \vec B (\vec x (\va,t),t) + \vec E (\vec x (\va,t),t) \right].
 \label{Lorflden}
 \enq
Now, since:
\beq
P \vec \nabla \cdot \vec \xi = \vec \nabla \cdot (P \vec \xi) - \vec \xi \cdot \vec \nabla P,
\label{Pxi}
 \enq
and using Gauss theorem the variation of the action can be written as:
\ber
\Delta {\cal A} &=&
\left. \int (\rho \vec v  + \rho_c \vec A)  \cdot \vec \xi dV \right|_{t1}^{t2}
 \nonumber \\
  &-& \int_{t1}^{t2}\left[\int (\rho \frac{d \vec v}{dt}- \vec f_{L} + \vec \nabla P)\cdot \vec \xi   dV  - \oint P \vec \xi \cdot d \vec \Sigma \right] dt.
 \label{varAfl4}
 \enr
 It follows that the variation of the action will vanish for a $\vec \xi$ such that $\vec \xi (t1) = \vec \xi (t2)=0$ and vanishing on a surface encapsulating the fluid, but other than that arbitrary only if the Euler equation for a charged fluid is satisfied, that is:
 \beq
 \frac{d \vec v}{dt}= -\frac{\vec \nabla P}{\rho} +\frac{\vec f_{L}}{\rho}
 \label{Eul1}
 \enq
 for the particular case that the fluid element is made of identical microscopic particles each
 with a mass $m$ and a charge $e$, it follows that the mass and charge densities are proportional to the number density $n$:
 \beq
 \rho = m~n, \quad \rho_c = e~n \Rightarrow \frac{\vec f_{L}}{\rho} = k \left[ \vec v \times \vec B + \vec E  \right],
 \label{Eul2}
 \enq
 thus except from the pressure terms the equation is similar to that of a point particle. In experimental fluid dynamics it is more convenient to describe a fluid in terms of quantities at
 a specific location, rather than quantities associated with unseen infinitesimal "fluid elements". This road leads to the Eulerian description of fluid dynamics and thinking in terms of flow fields rather than in terms of a velocity of "fluid elements" as will be discussed in the next section. This description will be shown later to be closely connected to quantum mechanics.

\section{An Eulerian Charged Fluid - the Clebsch Approach}

In this section we follows closely the analysis of \cite{Spflu} with the modification of taking into account the electromagnetic interaction which was neglected in our previous work. Consider the action:
\ber
 {\cal A} & \equiv & \int {\cal L} d^3 x dt, \qquad
{\cal L}  \equiv  {\cal L}_0 + {\cal L}_2 + {\cal L}_i
\nonumber \\
{\cal L}_0 & \equiv & \rho (\frac{1}{2} v^2 - \varepsilon), \qquad
{\cal L}_2 \equiv  \nu [\frac{\partial{\rho}}{\partial t} + \vec \nabla \cdot (\rho \vec v )]
- \rho \alpha \frac{d \beta}{dt}
 \nonumber \\
 {\cal L}_{i} &=&\rho_{c} \left(\vec A \cdot \vec v - \varphi \right)
\label{Lagactionsimpb}
\enr
In the Eulerian approach we consider the variational variables to be fields, that functions of
space and time. We have two such variational variables the vector velocity field $\vec v (\vec x,t)$ and density scalar field $\rho (\vec x,t)$. The conservation of quantities such as
the label of the fluid element, mass, charge and entropy are dealt by introducing Lagrange multipliers $\nu,\alpha$ in such a way that the variational principle will yield the following \ens:
\ber
& & \frac{\partial{\rho}}{\partial t} + \vec \nabla \cdot (\rho \vec v ) = 0
\nonumber \\
& & \frac{d \beta}{dt} = 0
\label{lagmulb}
\enr
Provided $\rho$ is not null those are just the continuity equation which ensures mass conservation and the conditions that $\beta$ is comoving and is thus a label, further analysis to be described below shows that $\alpha$ is a label too. This is why in the Eulerian approach we are obliged
to add the Lagrangian density ${\cal L}_2$. The specific internal energy $\varepsilon$ defined
in \ern{specifint} is dependent on the thermodynamic properties of the specific fluid. That is it generally depends through a given "equation of state" on the density and specific entropy. In our case we shall assume a barotropic fluid, that is a fluid in which $\varepsilon(\rho)$ is a function of the density $\rho$ only. Other functions connected to the electromagnetic interaction such as the potentials $\vec A, \varphi$ are assumed given function of coordinates and are not varied.
Another simplification which we introduce is the assumption the fluid element is made of microscopic
particles having a given mass $m$ and a charge $e$, in this case it follows from \ern{Eul2} that:
\beq
 \rho_c = k \rho.
 \label{rhoc}
 \enq

Let us take an arbitrary variational derivative of the above action with
respect to $\vec v$, this will result in:
\ber
\delta_{\vec v} A & = & \int d^3 x dt \rho \delta \vec v \cdot
[\vec v - \vec \nabla \nu - \alpha \vec \nabla \beta + k \vec A]
\nonumber \\
 & + & \oint d \vec S \cdot \delta \vec v \rho \nu + \int d \vec \Sigma \cdot \delta \vec v \rho [\nu],
\label{delActionvb}
\enr
the above boundary terms contain integration over the external boundary $\oint d \vec S$ and an
integral over the cut $\int d \vec \Sigma$ that must be introduced in case that $\nu$ is not single
valued, more on this case in later sections.
The external boundary term vanishes; in the case of astrophysical flows for which $\rho=0$ on the free flow
boundary, or the case in which the fluid is contained in a vessel which induces a no flux boundary condition $\delta \vec v \cdot \hat n =0$
($\hat n$ is a unit vector normal to the boundary). The cut "boundary" term vanish when the velocity field varies only parallel
to the cut that is it satisfies a Kutta type condition. If the boundary terms vanish  $\vec v$ must have the following form:
\beq
\vec v = \hat {\vec v} \equiv \alpha \vec \nabla \beta + \vec \nabla \nu -  k \vec A
\label{vformb}
\enq
this is a generalization of Clebsch representation of the flow field (see for example \cite{Eckart}, \cite[page 248]{Lamb H.}) for a charged flow. The vorticity of such a flow is:
\beq
\vec \omega = \vec \nabla \times \vec v =  \vec \nabla \alpha \times \vec \nabla \beta  -  k \vec B
\label{vortc1}
\enq
in which we have taken into account the definition of the magnetic field given in \ern{EB}.
Let us now take the variational derivative with respect to the density $\rho$, we obtain:
\ber
\delta_{\rho} A & = & \int d^3 x dt \delta \rho
[\frac{1}{2} \vec v^2 - w  - \frac{\partial{\nu}}{\partial t} -  \vec v \cdot \vec \nabla \nu+
k(\vec A \cdot \vec v - \varphi)]
\nonumber \\
 & + & \oint d \vec S \cdot \vec v \delta \rho  \nu +\int d \vec \Sigma \cdot \vec v \delta \rho  [\nu] +
  \int d^3 x \nu \delta \rho |^{t_1}_{t_0}
\label{delActionrhob}
\enr
in which $w = \frac{\partial(\rho \varepsilon )}{\partial \rho}$ is the specific enthalpy (see \ern{thermo2f}). Hence provided that $\delta \rho$ vanishes on the boundary of the domain, on the cut and in initial
and final times the following \eqn must be satisfied:
\beq
\frac{d \nu}{d t} = \frac{\partial \nu}{\partial t} + \vec v \cdot \vec \nabla \nu = \frac{1}{2} \vec v^2 - w + k(\vec A \cdot \vec v - \varphi)
\label{nueqb}
\enq
In the above we notice that taking a time derivative for a fixed label $\va$ (also known as a material derivative) of any quantity $g$ takes the form:
\beq
\frac{d g (\va,t) }{d t}  = \frac{d g (\vec x (\va,t),t) }{d t}
= \frac{\partial g}{\partial t} + \frac{d \vec x }{d t} \cdot \vec \nabla g
= \frac{\partial g}{\partial t} + \vec v \cdot \vec \nabla g
\label{materderiv}
\enq
once $g$ is considered to be a field dependent on $\vec x, t$. Finally we have to calculate the variation with respect to $\beta$, this will lead us to the following results:
\ber
\delta_{\beta} A & = & \int d^3 x dt \delta \beta
[\frac{\partial{(\rho \alpha)}}{\partial t} +  \vec \nabla \cdot (\rho \alpha \vec v)]
\nonumber \\
 & - & \oint d \vec S \cdot \vec v \rho \alpha \delta \beta -\int d \vec \Sigma \cdot \vec v \rho \alpha [\delta \beta]
 - \int d^3 x \rho \alpha \delta \beta |^{t_1}_{t_0}
\label{delActionchib}
\enr
Hence choosing $\delta \beta$ in such a way that the temporal and
spatial boundary terms vanish (this includes choosing $\delta \beta$ to be continuous on the cut if one needs
to introduce such a cut) in the above integral will lead to
the equation:
\beq
\frac{\partial{(\rho \alpha)}}{\partial t} +  \vec \nabla \cdot (\rho \alpha \vec v) =0
\enq
Using the continuity \ern{lagmulb} this will lead to the equation:
\beq
\frac{d \alpha}{dt} = 0
\label{alphacon}
\enq
Hence for $\rho \neq 0$ both $\alpha$ and $\beta$ are comoving coordinates.

\subsection{Euler's equations}
\label{Eulerequations}

We shall now show that a velocity field given by \ern{vformb}, such that the
functions $\alpha, \beta, \nu$ satisfy the corresponding equations
(\ref{lagmulb},\ref{nueqb},\ref{alphacon}) must satisfy Euler's equations.
Let us calculate the material derivative of $\vec v$:
\beq
\frac{d\vec v}{dt} = \frac{d\vec \nabla \nu}{dt}  + \frac{d\alpha}{dt} \vec \nabla \beta +
 \alpha \frac{d\vec \nabla \beta}{dt} - k \frac{d\vec A}{dt}
\label{dvform12b}
\enq
It can be easily shown that:
\ber
\frac{d\vec \nabla \nu}{dt} & = & \vec \nabla \frac{d \nu}{dt}- \vec \nabla v_n \frac{\partial \nu}{\partial x_n}
 = \vec \nabla (\frac{1}{2} \vec v^2 - w + k \vec A \cdot \vec v -k \varphi)- \vec \nabla v_n \frac{\partial \nu}{\partial x_n}
 \nonumber \\
 \frac{d\vec \nabla \beta}{dt} & = & \vec \nabla \frac{d \beta}{dt}- \vec \nabla v_n \frac{\partial \beta}{\partial x_n}
 = - \vec \nabla v_n \frac{\partial \beta}{\partial x_n}
  \label{dnablab}
\enr
In which $x_n$ is a Cartesian coordinate and a summation convention is assumed. Inserting the result from equations
(\ref{dnablab}) into \ern{dvform12b} yields:
\ber
& & \frac{d\vec v}{dt} = - \vec \nabla v_n (\frac{\partial \nu}{\partial x_n} +
 \alpha \frac{\partial \beta}{\partial x_n} ) + \vec \nabla (\frac{1}{2} \vec v^2 -w + k \vec A \cdot \vec v -k \varphi) - k \frac{d\vec A}{dt}
 \nonumber \\
&=& - \vec \nabla v_n (v_n + k A_n) + \vec \nabla (\frac{1}{2} \vec v^2 - w + k \vec A \cdot \vec v -k \varphi) - k \partial_t \vec A - k (\vec v \cdot \vec \nabla) \vec A
\nonumber \\
&=& - \vec \nabla w + k \vec E + k (v_n \vec \nabla A_n - v_n \partial_n \vec A),
\label{dvform2bb}
\enr
in the above we have used the electric field defined in \ern{EB}. We notice that
according to \ern{EB2}:
 \beq
 (v_n \vec \nabla A_n - v_n \partial_n \vec A)_l  = v_n (\partial_l  A_n -\partial_n  A_l)=
\epsilon_{lnj} v_n B_j = (\vec v \times \vec B)_l,
\label{EB3}
 \enq
 Hence we obtain the Euler equation of a charged fluid in the form:
 \beq
 \frac{d \vec v}{dt}= -\vec \nabla w + k \left[ \vec v \times \vec B + \vec E  \right]
 = - \frac{1}{\rho} \vec \nabla P + k \left[ \vec v \times \vec B + \vec E  \right],
  \label{Eul5}
 \enq
 since in a barotropic fluid (see \ern{thermo2g}):
 \beq
 \vec \nabla w = \frac{\partial w }{\partial \rho} \vec \nabla \rho =
  \frac{1}{\rho} \frac{\partial P }{\partial \rho} \vec \nabla \rho = \frac{1}{\rho} \vec \nabla P.
  \label{Eul6}
 \enq
  The above equation is identical to \ern{Eul1} and thus proves that the Euler equations can be derived from the action given in \ern{Lagactionsimpb} and hence
all the equations of charged fluid dynamics can be derived from the above action
without restricting the variations in any way.

\subsection{Simplified action}
\label{simpact}
The reader of this paper might argue that the authors have introduced unnecessary complications
to the theory of fluid dynamics by adding three  more functions $\alpha,\beta,\nu$ to the standard set
$\vec v,\rho$. In the following we will show that this is not so and the action given in \ern{Lagactionsimpb} in
a form suitable for a pedagogic presentation can indeed be simplified. It is easy to show
that the Lagrangian density appearing in \ern{Lagactionsimpb} can be written in the form:
\ber
{\cal L} & = & -\rho [\frac{\partial{\nu}}{\partial t} + \alpha \frac{\partial{\beta}}{\partial t}
+\varepsilon (\rho) +k \varphi] +
\frac{1}{2}\rho [(\vec v-\hat{\vec v})^2-\hat{\vec v}^2]
\nonumber \\
& + &  \frac{\partial{(\nu \rho)}}{\partial t} + \vec \nabla \cdot (\nu \rho \vec v )
\label{Lagactionsimpb4}
\enr
In which $\hat{\vec v}$ is a shorthand notation for
$\vec \nabla \nu + \alpha \vec \nabla \beta -k \vec A $
(see \ern{vformb}). Thus ${\cal L}$ has three contributions:
\ber
{\cal L} & = & \hat {\cal L} + {\cal L}_{\vec v}+ {\cal L}_{boundary}
\nonumber \\
\hat {\cal L} &\equiv & -\rho [\frac{\partial{\nu}}{\partial t} + \alpha \frac{\partial{\beta}}{\partial t}
+\varepsilon (\rho)+ k \varphi + \frac{1}{2}(\vec \nabla \nu + \alpha \vec \nabla \beta -k \vec A )^2]
\nonumber \\
{\cal L}_{\vec v} &\equiv & \frac{1}{2}\rho (\vec v-\hat{\vec v})^2
\nonumber \\
{\cal L}_{boundary} &\equiv & \frac{\partial{(\nu \rho)}}{\partial t} + \vec \nabla \cdot (\nu \rho \vec v )
\label{Lagactionsimp5b}
\enr
The only term containing $\vec v$ is ${\cal L}_{\vec v}$, it can easily be seen that
this term will lead, after we nullify the variational derivative, to \ern{vformb} but will otherwise
have no contribution to other variational derivatives. Notice that the term ${\cal L}_{boundary}$
contains only complete partial derivatives and thus can not contribute to the equations although
it can change the boundary conditions. Hence we see that \ers{lagmulb}, \ern{nueqb} and \ern{alphacon}
can be derived using the Lagrangian density $\hat {\cal L}$ in which $\hat{\vec v}$ replaces
$\vec v$ in the relevant equations. Furthermore, after integrating the four \eqs
(\ref{lagmulb},\ref{nueqb},\ref{alphacon}) we can insert the potentials $\alpha,\beta,\nu$
into \ern{vformb} to obtain the physical velocity $\vec v$.
Hence, the general barotropic fluid dynamics problem is changed such that instead of
solving the Euler and continuity equations we need to solve an alternative set which can be
derived from the Lagrangian density $\hat {\cal L}$.

The Lagrangian density $\hat {\cal L}$ has two manifest properties that will repeat in both
Schrodinger and Pauli quantum mechanics, but are not evident in other formulations of classical mechanics. First it contains a term quadratic in the vector potential $\vec A$ this cannot be found
in classical Lagrangians which are described in previous sections, in those cases the Lagrangians are always linear in $\vec A$. Second the gauge freedom described in \ern{gauge} is preserved
by redefining the variation variable:
\beq
\nu' =\nu - k \Lambda.
\label{gauge2b}
\enq
However, before we discuss quantum theory a remark on Fisher information is required.

\section{Fisher Information}

Let there be a random variable $X$ with probability density function (PDF) $f_X (x)$.
The Fisher Information for a PDF which is translationaly invariant is given by the form:
\beq
F_I = \int dx {\left(\frac{df_X}{dx}\right)}^2 \frac{1}{f_X}
\label{Fisher}
\enq
It was shown \cite{Fisher,Frieden} that the standard deviation $\sigma_X$ of any random variable is bounded
from below such that:
\beq
\sigma_X \ge \sigma_{X min} = \frac{1}{F_I}
\label{Fisherb}
\enq
Hence the higher the Fisher information we have about the variable the smaller standard deviation we may achieve and thus our
knowledge about the value of this random variable is greater. This is known as the Cramer Rao inequality. Fisher information is most elegantly introduced in terms of
the probability amplitude:
\beq
f_X = a^2 \Rightarrow F_I = 4\int dx {\left(\frac{da}{dx}\right)}^2
\label{Fisher2}
\enq
In this work we will be interested in a three dimensional random variable designating the position of an electron, hence:
\beq
F_I = \int d^3 x {\left(\vec \nabla f_{\vec X} \right)}^2 \frac{1}{f_{\vec X}} = 4 \int d^3 x {\left(\vec \nabla a \right)}^2 \equiv \int d^3 x {\cal F}_I
\label{Fisher3}
\enq
In the above ${\cal F}_I \equiv 4 {\left(\vec \nabla a \right)}^2$ is the Fisher information density. The Fisher information is the only intuitive reason to consider a "probability density
amplitude" a notion which is quite important in quantum mechanics as it is also the amplitude of
the quantum wave function to be described in the next section.

\section {Schr\"{o}dinger's Theory Formulated in Terms of Fluid Mechanics}

\subsection {Background}

Quantum mechanics has lost faith in our ability to predict precisely the whereabouts of even a single particle. What the theory does predict precisely is the evolution in time of a quantity
denoted "the quantum wave function", which is related to a quantum particle whereabouts in a statistical manner. This evolution is described by an equation suggested by
Schr\"{o}dinger \cite{Schrodinger}:
\beq
i \hbar \dot{\psi} = \hat{H} \psi, \qquad \hat{H} = -\frac{1}{2m}
 \left(\hbar \vec \nabla - i e \vec A\right)^2 + e \varphi
 \label{Seq}
 \enq
in the above $i=\sqrt{-1}$ and $\psi$ is the complex wave function. $\dot{\psi}=\frac{\partial \psi }{\partial t}$ is the partial time
derivative of the wave function. $\hbar=\frac{h}{2 \pi}$ is Planck's constant divided by $2 \pi$  and $m$ is the particles mass.
The Lagrangian density ${\cal L}$ for the non-relativistic electron is written as:
\beq
 {\cal L}_S=  \frac{1}{2} i \hbar (\psi^* \dot{\psi} - \dot{\psi^* } \psi ) - \psi^* \hat{H} \psi
 \label{nrLag}
 \enq
 A strait forward variation of the action:
 \beq
 {\cal A}_S \equiv \int_{t1}^{t2} L_S dt \equiv \int_{t1}^{t2} \int {\cal L}_S d^3 x
 \label{ACS}
 \enq
 with respect to $\psi^*$ will lead to \ern{Seq} (while a variation with respect to $\psi$
 will lead to a complex conjugate of the same). However, this presentation of quantum mechanics is
 rather abstract and does not give any physical picture regarding the meaning of the quantities involved. Thus we write the quantum wave function using its modulus $a$  and phase $\phi$:
 \beq
\psi = a e^{i\phi }
\label{psi1}
\enq
 the Lagrangian density takes the form:
\beq
{\cal L}_S= -\frac{\hbar^2}{2m} [(\vec \nabla a)^{2} +a^{2}(\vec \nabla \phi)^{2}] - e a^{2} \varphi   - \hbar a^{2} \frac{\partial \phi }{\partial t} + e a^2 \vec A \cdot \vec v_S +
 \frac{e^2}{2m} a^2 A^2
\label{nrLag2}
\enq
in we define:
\beq
\vec v_S = \frac{\hbar}{m} \vec \nabla \phi - \frac{e}{m} \vec A
\label{vS}
\enq
The variational derivative of this with respect to $\phi$ yields
the \ce:
 \beq
\frac{\delta {\cal A}_S}{\delta \phi}  = 0 \rightarrow
 \frac{ \partial\hat \rho }{ \partial t} + \vec \nabla \cdot (\hat \rho \vec v_S) = 0
 \label{nrec}
 \enq
 in which the mass density is defined as:
 \beq
 \hat \rho= m a^2.
 \label{massden}
 \enq
 Hence $\vec v_S$ field is the velocity associated with the probability, charge and mass flow. Variationally deriving with respect to $a$ leads to the Hamilton Jacobi equation:
 \beq
 \frac{\delta  {\cal A}_S}{\delta a} = 0 \rightarrow
 \frac{\partial S}{\partial t} +
  \frac{1}{2m}\left(\vec \nabla S -e \vec A\right)^2 +  e \varphi =  \frac{\hbar^2 \nabla^2 a }{ 2 m a}
 \label{nrhje}
 \enq
 in which: $S = \hbar \phi$. The right hand side of the above equation
 contains the "quantum correction".
 These results are elementary, but their derivation illustrates the advantages
 of using the two variables, phase and modulus, to obtain equations of motion
that have a  substantially different form than the familiar \SE (although having the same
mathematical content) and have  straightforward physical interpretations \cite{Bohm}. The interpretation is,
 of course, connected to the modulus being a physical observable (by Born's
 interpretational postulate) and to the phase having a similar though somewhat
 more problematic status. (The "observability" of the phase has been
 discussed in the literature by various sources, e.g. in \cite{Mandel} and,
 in connection with a recent development, in \cite{EY1,EY2}.)

\subsection{Similarities Between Potential Fluid Dynamics and Qua\-ntum
 Mechanics}

In writing the Lagrangian density  of quantum mechanics in the modulus-phase
representation,
\er{nrLag2}, one notices a striking similarity between this Lagrangian density
 and that
of potential fluid dynamics (fluid dynamics without vorticity) as represented in
the work of Clebsch \cite{Seliger}.
The connection between fluid dynamics and quantum mechanics
of an electron was already discussed by Madelung \cite{Madelung} and in Holland's book \cite{Holland}.
However, the discussion by Madelung refers to the equations only and does not
address the variational formalism which was discussed in \cite{Spflu} and is repeated here with the important addition of an electromagnetic interaction.

If a flow satisfies the condition of zero vorticity in the absence of a magnetic field, i.e. the velocity field $\vec v$ is such that $\vec \nabla \times \vec v = 0$, then there exists a function $\nu$ such that $\vec v = \vec \nabla \nu$.
The above statement is equivalent to taking  a Clebsch representation  of the velocity field but with $\alpha=\beta=0$.
In that case following \ern{Lagactionsimp5b} one can describe the fluid mechanical system with the following Lagrangian density:
\beq
\hat {\cal L} = -[ \frac{\partial \nu}{\partial t}  + \frac{1}{2} (\vec \nabla \nu - k \vec A)^2 + \varepsilon(\rho) + k \varphi] \rho
\label{Asheractpc}
\enq
by inserting $\alpha=\beta=0$ in $\hat {\cal L}$ \cite{Fisherspin}.
Taking the variational derivative with respect to $\nu$ and $\rho$, one obtains the following equations:
\ber
\frac {\partial \rho }{\partial t}  & + & \vec \nabla \cdot (\rho (\vec \nabla \nu - k \vec A)) = 0
\label{ContinS2p}
 \\
\frac {\partial \nu}{\partial t} & = & -\frac{1}{2} (\vec \nabla \nu - k \vec A)^2 - w - k \varphi.
\label{nueq2p}
\enr
The first of those \eqs is the continuity equation, while the second is Bernoulli's
equation.

Going back to the quantum mechanical system described by \ern{nrLag2}, we
introduce the following variable: $ \hat \nu = \frac{\hbar \phi}{m} = \frac{S}{m}$.
In terms of these new variables the Lagrangian density in \ern{nrLag2} will take
the form:
\beq
{\cal L}_S = - [ \frac{\partial \hat \nu}{\partial t}  +
\frac{1}{2} (\vec \nabla \hat \nu - k \vec A)^2 +
  \frac{\hbar^2}{2 m^2} \frac{(\vec{\nabla}\sqrt{\hat \rho})^2}{\hat \rho}  + k \varphi ] \hat \rho
\label{fluidquantum}
\enq
When compared with \ern{Asheractpc} the following correspondence is noted:
\beq
 \hat \nu \Leftrightarrow \nu, \qquad  \hat \rho \Leftrightarrow \rho, \qquad
 \frac{\hbar^2}{2 m^2} \frac{(\vec{\nabla}\sqrt{\hat \rho})^2}{\hat \rho}
\Leftrightarrow \varepsilon.
\label{corres}
\enq
The quantum "internal energy":
\beq
\varepsilon_q \equiv  \frac{\hbar^2}{2 m^2} \frac{(\vec{\nabla}\sqrt{\hat \rho})^2}{\hat \rho}
\label{intenq}
\enq
depends also on the derivative of the density and in this sense it is non local. This is unlike
the fluid case, in which internal energy is a function of the mass density only. However, in both cases the internal energy is a positive quantity.
Unlike classical systems in which the Lagrangian is quadratic in the time
derivatives of the degrees of freedom (see \ern{classparticleact}), the Lagrangians of both quantum and Eulerian fluid dynamics are linear in the time derivatives of the degrees of freedom. We also
note that gauge freedom is preserved (see \ern{gauge}) provided that the variational
quantum potential $\hat \nu$  (or quantum phase) is redefined as:
\beq
\hat \nu' =\hat \nu - k \Lambda, \qquad \phi' =\phi - \frac{e}{\hbar} \Lambda,
\label{gauge2b2}
\enq
this means that the global phase of a quantum wave function does not have a physical meaning,
in contrast to the quantity $\vec v_S$ that does (see \ern{vS}), as it is invariant under gauge transformations.
Finally we note that the concept of quantum internal energy is closely related to its variational derivative the concept of quantum potential \cite{Holland} (see the right side of \ern{nrhje}):
\beq
Q = -\frac{\hbar^2}{2 m} \frac{\vec{\nabla}^2 \sqrt{\hat \rho}}{\sqrt{\hat \rho}}.
\label{qupo}
\enq
And also that in the limit $\hbar\rightarrow 0$ \Sc's quantum mechanics is essentially a potential
fluid flow without pressure or internal energy.

\subsection{Madelung flows in terms of Fisher information}

As explained in the introduction the quantum Madelung flow does not have a microstructure that will explain its internal energy.
To understand the origins of this term let us look at the internal energy of \ern{fluidquantum},
the term appearing in the Lagrangian density has the form:
\beq
\hat \rho \varepsilon_q = \frac{\hbar^2}{2 m^2} (\vec{\nabla}\sqrt{\hat \rho})^2 = \frac{\hbar^2}{2 m} (\vec \nabla a)^2.
\label{inte}
\enq
Comparing this to \ern{Fisher3} we arrive at the result:
\beq
\hat \rho \varepsilon_q  = \frac{\hbar^2}{8 m} {\cal F}_{Iq}. \qquad {\cal F}_{Iq} \equiv 4 (\vec \nabla a)^2
\label{inte2}
\enq
Thus the Lagrangian density of the Madelung flow can be written as:
\beq
{\cal L} = - [ \frac{\partial \hat \nu}{\partial t}  +
 \frac{1}{2} (\vec \nabla \hat \nu - k \vec A)^2 +k \varphi] \hat \rho -
\frac{\hbar^2}{8 m} {\cal F}_{Iq}
\label{fluidquantumf}
\enq
The pre-factor $\frac{\hbar^2}{8 m}$ seems to appear in every case in which Fisher information appears in a quantum Lagrangian
and may be significant. This is the case also in spin fluid dynamics as will be shown in the next section.

Thus Schr\"{o}dinger's fluid differ from a classical potential flow in that it takes into account Fisher information as a driving force, and its "fluid element" lacks an internal energy, that is, it is truly microscopic and has no structure. Thus when written explicitly in terms of physical quantities and not mere abstractions, quantum mechanics reveals itself as a theory that takes into account information gain (or loss) as a force of nature. We conclude this section by quoting Anton Zeilinger's recent remark to the press, that it is quantum mechanics that demonstrates that information is more fundamental than space-time.

\section {Spin}

\Sc's quantum mechanics is limited to the description of spin less particles and its fluid dynamics representation is limited
to zero vorticity (for  the zero magnetic field case) potential flows. This suggests that a quantum theory of particles with spin may have a fluid dynamics representation
which cannot be described solely by using $\hat \nu$ and requires the full Clebsch apparatus. The Pauli equation for a non-relativistic particle with spin is given by:
\beq
i \hbar \dot{\psi} = \hat{H} \psi, \qquad \hat{H} = -\frac{\hbar^2}{2m}[\vec \nabla-\frac{ie}{\hbar }\vec A]^2 +
\mu \vec B \cdot \vec \sigma + e \varphi
 \label{Pauli}
 \enq
$\psi$ here is a two dimensional complex column vector (also denoted as spinor), $\hat{H}$ is a two dimensional hermitian operator matrix, $\mu$ is the magnetic moment of the particle. $\vec \sigma$ is a vector of two dimensional Pauli matrices which can be represented as follows:
\beq
\sigma_{1} = \left( \begin{array}{cc} 0 & 1 \\ 1 & 0 \end{array} \right), \qquad
\sigma_{2} = \left( \begin{array}{cc} 0 & -i \\ i & 0 \end{array} \right), \qquad
\sigma_{3} = \left( \begin{array}{cc} 1 & 0 \\ 0 & -1 \end{array} \right)
\label{sigma}
\enq
A spinor $\psi$ satisfying \ern{Pauli} must also satisfy a continuity equation of the form:
\beq
\frac{\partial{\rho}}{\partial t} + \vec \nabla \cdot  \vec j  = 0.
\label{massconp}
\enq
In the above:
\beq
\rho = \psi^\dagger \psi, \qquad \vec j =  \frac{\hbar}{2mi} [\psi^\dagger \vec \nabla \psi -(\vec \nabla \psi^\dagger) \psi]
- k \vec{A}\rho.
\label{pcd}
\enq
The symbol $\psi^\dagger$ represents a row spinor (the transpose) whose components are equal to the complex conjugate of
the column spinor $\psi$. Comparing the standard continuity equation to \ern{massconp} suggests the definition of a velocity field as follows \cite{Holland}:
\beq
\vec v = \frac{\vec j}{\rho}= \frac{\hbar}{2mi\rho} [\psi^\dagger \vec \nabla \psi -(\vec \nabla \psi^\dagger) \psi] - k \vec{A}.
\label{pv}
\enq
A variational description of the Pauli system can be given using the following Lagrangian density:
\beq
 {\cal L}=  \frac{1}{2} i \hbar (\psi^\dagger \dot{\psi} - \dot{\psi^\dagger } \psi )  - \psi^\dagger \hat H \psi
 \label{PLag}
 \enq
 Holland \cite{Holland} has suggested the following representation of the spinor:
 \beq
\psi = R e^{i\frac{\chi}{2}} \left( \begin{array}{c} \cos \left(\frac{\theta}{2}\right) e^{i\frac{\phi}{2}} \\
i \sin \left(\frac{\theta}{2}\right) e^{-i\frac{\phi}{2}}  \end{array} \right) \equiv
\left( \begin{array}{c} \psi_{\uparrow} \\ \psi_{\downarrow} \end{array} \right).
\label{psiH}
\enq
In terms of this representation the density is given as:
\beq
\rho = \psi^\dagger \psi = R^2 \Rightarrow R= \sqrt{\rho}.
\label{rhopsi}
\enq
The mass density is given as:
\beq
\hat \rho = m \rho = m \psi^\dagger \psi =m R^2.
\label{rhom}
\enq
The probability amplitudes for spin up and spin down electrons are given by:
\beq
a_{\uparrow} = \left | \psi_{\uparrow} \right| = R \left | \cos \frac{\theta}{2} \right|, \qquad
a_{\downarrow} = \left | \psi_{\downarrow} \right| = R \left | \sin \frac{\theta}{2} \right|
\label{probamp}
\enq
Let us now look at the expectation value of the spin:
\beq
<\frac{\hbar}{2} \vec \sigma> = \frac{\hbar}{2}\int \psi^\dagger \vec \sigma \psi d^3 x =
 \frac{\hbar}{2}\int \left(\frac{\psi^\dagger \vec \sigma \psi}{\rho}\right) \rho d^3 x
\label{spinex}
\enq
The spin density can be calculated using the representation given in \ern{psiH} as:
\beq
\hat s \equiv \frac{\psi^\dagger \vec \sigma \psi}{\rho} = (\sin \theta \sin \phi, \sin \theta \cos \phi, \cos \theta)
\label{spinden}
\enq
This gives an easy physical interpretation to the variables $\theta,\phi$ as angles which describe the projection
of the spin density on the axes. $\theta$ is the elevation angle of the spin density vector and $\phi$ is
the azimuthal angle of the same. The velocity field can now be calculated by inserting $\psi$ given in \ern{psiH}
into \ern{pv}:
\beq
\vec v = \frac{\hbar}{2m} (\vec \nabla \chi + \cos \theta \vec \nabla \phi) - k \vec{A}.
\label{pv2}
\enq
Comparing \ern{pv2} with the generalized Clebsch form given in \ern{vformb} suggest the following identification:
\beq
\alpha=\cos \theta, \qquad \beta=\frac{\hbar}{2m} \phi, \qquad \nu = \frac{\hbar}{2m} \chi.
\label{pclebsch}
\enq
Notice that $\alpha$ is single valued, but $\beta$ and $\nu$ are not.
Obviously this velocity field will have a generically non vanishing vorticity even if the magnetic field is null:
\ber
\vec \omega &=& \vec \nabla \times \vec v = \vec \nabla \alpha \times \vec \nabla \beta - k \vec B
\nonumber \\
&=& \frac{\hbar}{2m} \vec \nabla \cos \theta \times \vec \nabla \phi - k \vec B
= \frac{\hbar}{2m} \sin \theta \vec \nabla \phi \times \vec \nabla \theta - k \vec B.
\label{vorticp}
\enr
we see again that the gauge symmetry  (\ern{gauge}) is maintained provided that the variational
variable $\nu$ proportional to the global phase $\chi$ is redefined:
\beq
\nu' =\nu - k \Lambda.
\label{gauge2bb}
\enq
Inserting the representation of $\psi$ given in \ern{psiH} into the Lagrangian density \ern{PLag} will yield after tedious
but straight forward calculations the Lagrangian density:
\ber
{\cal L}_P &\equiv& - \hat \rho [\frac{\partial{\nu}}{\partial t} + \alpha \frac{\partial{\beta}}{\partial t}
+\varepsilon_{qt} +k \varphi+ \frac{1}{2}(\vec \nabla \nu + \alpha \vec \nabla \beta- k \vec A )^2
+ \frac{\mu}{m} \vec B \cdot \hat s]
\nonumber \\
\varepsilon_{qt}[\hat \rho,\alpha,\beta] &\equiv & \varepsilon_q [\hat \rho] + \varepsilon_{qs}[\alpha,\beta]
\nonumber \\
\varepsilon_q [\hat \rho] &\equiv & \frac{\hbar^2}{2 m^2} \frac{(\vec{\nabla}\sqrt{\hat \rho})^2}{\hat \rho} =
 \frac{\hbar^2}{2 m^2} \frac{(\vec{\nabla}R)^2}{R^2}
 \nonumber \\
\varepsilon_{qs} [\alpha,\beta] &\equiv & \frac{\hbar^2}{8 m^2} \left((\vec \nabla \theta)^2 + \sin^2 \theta (\vec \nabla \phi)^2 \right)
\nonumber \\
&=& \frac{1}{2} \left(\left(\frac{\hbar}{2m}\right)^2\frac{(\vec \nabla \alpha)^2}{1-\alpha^2} + (1-\alpha^2) (\vec \nabla \beta)^2 \right)
\label{LagdP}
\enr
The Lagrangian ${\cal L}_P$ has the same form as the Clebsch Lagrangian $\hat {\cal L}$ given in \ern{Lagactionsimp5b}. However, there are some important differences. The internal energy in the Pauli Lagrangian is positive as for the barotropic
fluid but now the internal energy depends on the derivatives of the degrees of freedom and not just on the density at given point
in this sense this internal energy is non local. Moreover, it is made of two part the \Sc \ quantum internal energy $\varepsilon_q$ which depends on the mass density and the spin quantum internal energy $\varepsilon_{qs}$ that depend on the spin (vorticity) degrees of
freedom. Finally the classical limit $\hbar\rightarrow 0$ will eliminate $\varepsilon_q$ but will not eliminate the spin internal
energy:
\beq
\lim_{\hbar\rightarrow 0} \varepsilon_{qs} = \frac{1}{2} (1-\alpha^2) (\vec \nabla \beta)^2
\enq
In this sense the Pauli theory has no standard classical limit, although this limit is a perfectly legitimate classical field theory. The interaction terms of electron dipole moment
with the magnetic field  $\frac{\mu}{m} \vec B \cdot \hat s$ has no classical analogue either. Indeed, this term was introduced by Pauli on a empirical basis with no theoretical justification, that is Pauli was aiming to explain the Stern-Gerlach experiment. Later, however, it was obtained from the relativistic Dirac equation. This points to the conclusion that we have gone as far as possible in understanding quantum mechanics is a non-relativistic framework. And thus further insight can only be gained by considering relativistic fluids, which are unfortunately beyond the scope of the current paper. This is indeed peculiar, as sometimes it is claimed that quantum mechanics and relativity (mainly general relativity) are contradictory when it now seems to be that quantum mechanics cannot be understood without relativity.

We conclude this section by writing down the equations for the Clebsch-Pauli variables,
unfortunately those equations are not very helpful. Taking the variational derivative we arrive at the equations of motion (see also \cite{Holland} p. 392 equations (9.3.6), (9.3.17), (9.3.18)) :
\ber
& & \frac{\partial{\hat \rho}}{\partial t} + \vec \nabla \cdot ( \hat \rho \vec v ) = 0
\nonumber \\
& & \frac{d \alpha}{d t} = \frac{1}{\hat \rho} \vec \nabla \cdot \left( \hat \rho (\alpha^2-1)\vec \nabla \beta \right) + \frac{2 \mu}{\hbar}(\vec B \times \hat s)_3
\nonumber \\
& & \frac{d \beta}{d t} = \left(\frac{\hbar}{2m}\right)^2 \frac{1}{\hat{\rho} \sqrt{1-\alpha^2}} \vec \nabla \cdot \left(\hat{\rho} \frac{\vec \nabla \alpha}{\sqrt{1-\alpha^2}}\right) + \alpha (\vec \nabla \beta)^2
\nonumber \\
&+& \frac{\mu}{m}\vec B \cdot (\cot \theta \sin \Phi, \cot \theta \cos \Phi, -1)
\nonumber \\
& & \frac{d{\nu}}{d t} = \frac{1}{2} \vec v^2 - e \varphi - \alpha^2 (\vec \nabla \beta)^2 - \frac{Q}{m}-\varepsilon_{qs}
\nonumber \\
& & -\left(\frac{\hbar}{2m}\right)^2 \frac{\alpha}{\hat{\rho} \sqrt{1-\alpha^2}} \vec \nabla \cdot \left(\hat{\rho} \frac{\vec \nabla \alpha}{\sqrt{1-\alpha^2}}\right) -\frac{\mu}{m} \vec B \cdot \hat s.
\label{spineq}
\enr
We notice that in spin fluid dynamics $\alpha$ and $\beta$ are not comoving scalar fields (labels) as in the case of ideal barotropic fluid dynamics. We are now in a position to calculate the material derivative of the velocity and obtain the spin fluid dynamics Euler equation
(\cite{Holland} p. 393 equation (9.3.19)):
\beq
\frac{d \vec v}{d t}  =- \vec \nabla (k \varphi + \frac{Q}{m})-\left(\frac{\hbar}{2m}\right)^2
\frac{1}{\hat \rho} \partial_k(\hat \rho \vec \nabla \hat s_j \partial_k \hat s_j)
+k (\vec E + \vec v \times \vec B) - \frac{\mu}{m} (\vec \nabla B_j) s_j
\label{EulerP}
\enq

\subsection{Spin flows in terms of Fisher information}

As explained in the introduction the quantum Spin flow does not have a microstructure that will explain its internal energy, so it should not have an internal energy. Nevertheless, the Lagrangian density does contain a "quantum internal energy term" as can be seen in \ern{LagdP}.
The said term can be written as:
\beq
\hat \rho \varepsilon_{qt} = \hat \rho \varepsilon_{q}+\hat \rho \varepsilon_{qs}
= \frac{\hbar^2}{8 m} \left[ 4 (\vec \nabla R)^2 + R^2 (\vec \nabla \theta)^2 + R^2 \sin^2 \theta (\vec \nabla \phi)^2 \right]
\label{intes}
\enq
Using the amplitudes of \ern{probamp} and the definition of Fisher information density of \ern{Fisher3} we arrive at the result:
\beq
{\cal F}_{Ip} = {\cal F}_{I \uparrow} + {\cal F}_{I \downarrow} = 4 \left[(\vec \nabla a_{\uparrow})^2 + (\vec \nabla a_{\downarrow})^2\right]
= 4 (\vec \nabla R)^2 + R^2 (\vec \nabla \theta)^2
\label{fishers}
\enq
Hence:
\beq
\hat \rho \varepsilon_{qt} = \frac{\hbar^2}{8 m} {\cal F}_{Ip}+ \frac{1}{2} \hat \rho (1-\alpha^2) (\vec \nabla \beta)^2
\label{inte2s}
\enq
Thus the Lagrangian density of the spin flow  given in \ern{LagdP} can be written as:
\beq
{\cal L}_P =
- \hat \rho [\frac{\partial{\nu}}{\partial t} + \alpha \frac{\partial{\beta}}{\partial t}
+\frac{1}{2} (1-\alpha^2) (\vec \nabla \beta)^2 +k \varphi+ \frac{1}{2}(\vec \nabla \nu + \alpha \vec \nabla \beta - k \vec A)^2]
-\frac{\hbar^2}{8 m} {\cal F}_{Ip}
\label{LagdPs}
\enq
The pre-factor $\frac{\hbar^2}{8 m}$ seems to appear in every case in which Fisher information appears in a quantum Lagrangian
and may be significant.

\section {Conclusion}

In this work which is a continuation of previous papers \cite{Spflu,Fisherspin,Fisherspin2} it was shown how Pauli's theory can be formulated as a spin fluid dynamics it terms of a generalized Clebsch representation modified to take into account electromagnetic field which may affect a charged fluid.
The theory is given in terms of a variational principle and the fluid equations are derived.
The similarities and differences with barotropic fluid dynamics were discussed.

 A fundamental problem in the fluid mechanical interpretation of quantum mechanics still exist. This refers to the meaning of thermodynamic quantities which are part of fluid mechanics, and imply that a fluid element is not a point particle but has internal structure and thus internal energy. In thermodynamics concepts like specific enthalpy,  pressure and temperature are
 derivatives of the specific internal energy which is given in terms of the equation of state as function of entropy and density.
 The internal energy is a part of any Lagrangian density attempting to describe fluid dynamics.
 The form of the internal energy can in principle be explained on the basis of the microscopic composition of the fluid, that is the atoms and
 molecules from which the fluid is composed and their interactions using statistical mechanics. However, the quantum fluid has no microscopic structure
 and yet analysis of the equations of both the spin less \cite{Madelung,Complex} and spin \cite{Spflu} quantum fluid dynamics shows that terms analogue
 to internal energies appear in both cases. The question then arises where do those internal energies come from, surely one would not suggest that the
 quantum fluid has a microscopic sub structure as this will defy the conception of the electron as a fundamental particle. The answer to this question
comes from an entirely different discipline of measurement theory \cite{Fisher}. Fisher information a basic notion of measurement theory  is a measure of the quality of the  measurement of any quantity. It was shown that this concept is proportional to the internal energy of a spin less electron and Schr\"{o}dingers theory is essentially a theory of a potential flow which moves under the
influence of electromagnetic and fisher information forces. Fisher information
 can also explain  most parts of the internal energy of an electron with spin. This puts
(Fisher) information as a fundamental force of  nature, which has the same status as electromagnetic forces in the quantum mechanical level of reality. Indeed according to  Anton Zeilinger's recent remark to the press, it is quantum mechanics that demonstrates that information is more fundamental than space-time.

We have highlighted the similarities between the variational principles of Eulerian fluid mechanics and both Schr\"{o}dinger's and Pauli's quantum mechanics as opposed to classical mechanics. The former have only linear time derivatives of degrees of freedom while that later have quadratic time derivatives. The former contain terms quadratic in the vector potential $\vec A$ while the later contain only linear terms, thus making the mass current and the current which couple to vector potential proportional. The former are manifestly gauge invariant which is achieved by redefining a
phase or a flow potential.

  To conclude we suggest the following future directions of research:
\begin{enumerate}
  \item It is conjectured that same analogy found between Pauli's theory and fluid dynamics may be found between  Dirac's relativistic electron theory and relativistic fluid dynamics.
   \item It is also conjectured that an Eulerian relativistic fluid once properly defined in terms of a variational principle and augmented by a four dimensional Fisher information term will fully explain Dirac's relativistic electron theory and inter alia Pauli's theory as well.
\end{enumerate}


\begin {thebibliography}9

 \bibitem{Kant}
Kant, I. (1781). Critik der reinen Vernunft.
\bibitem{Bohm} D. Bohm, {\it Quantum Theory} (Prentice Hall, New York, 1966)  section 12.6
 \bibitem{Holland}
P.R. Holland {\it The Quantum Theory of Motion} (Cambridge University Press, Cambridge, 1993)
 \bibitem{DuTe}
D. Durr \& S. Teufel {\it Bohmian Mechanics: The Physics and Mathematics of Quantum Theory} (Springer-Verlag, Berlin Heidelberg, 2009)
\bibitem{Madelung}
E. Madelung, Z. Phys., {\bf 40} 322 (1926)
\bibitem {Complex}
R. Englman and A. Yahalom "Complex States of Simple Molecular Systems"
a chapter of the volume "The Role of Degenerate States in Chemistry" edited by M.
Baer and G. Billing in Adv. Chem. Phys. Vol. 124 (John Wiley \& Sons 2002). [Los-Alamos Archives physics/0406149]
\bibitem{Pauli}
W. Pauli (1927) Zur Quantenmechanik des magnetischen Elektrons Zeitschrift f\"{u}r Physik (43) 601-623
\bibitem {EYB1}
R. Englman, A.Yahalom and M. Baer, J. Chem. Phys.{109} 6550 (1998)
\bibitem {EYB2}
R. Englman, A. Yahalom and M. Baer, Phys. Lett. A {\bf 251} 223 (1999)
\bibitem {EY1}
R. Englman and A. Yahalom, Phys. Rev. A {\bf 60} 1802 (1999)
\bibitem {EYB3}
R. Englman, A.Yahalom and M. Baer, Eur. Phys. J. D {\bf 8} 1 (2000)
\bibitem {EY2}
R. Englman and A.Yahalom, Phys. Rev. B {\bf 61} 2716 (2000)
\bibitem {EY3}
R. Englman and A.Yahalom, Found. Phys. Lett. {\bf 13} 329 (2000)
\bibitem {EY4}
R. Englman and A.Yahalom, {\it The Jahn Teller Effect: A Permanent Presence
in the Frontiers of Science} in M.D. Kaplan and G. Zimmerman (editors),
{\it Proceedings of the NATO Advanced Research
Workshop, Boston, Sep. 2000}  (Kluwer, Dordrecht, 2001)
\bibitem {BE2}
M. Baer and R. Englman, Chem. Phys. Lett. {\bf 335} 85 (2001)
\bibitem {MBEY}
A. Mebel, M. Baer, R. Englman and A. Yahalom, J.Chem. Phys. {\bf 115} 3673 (2001)
\bibitem {EYBM4MCI}
R. Englman \& A. Yahalom, "Signed Phases and Fields Associated with Degeneracies" Acta Phys. et Chim., 34-35, 283 (2002). [Los-Alamos Archives - quant-ph/0406194]
\bibitem {EY5}
R. Englman, A. Yahalom and M. Baer,"Hierarchical Construction of Finite Diabatic Sets, By Mathieu Functions", Int. J. Q. Chemistry, 90, 266-272 (2002). [Los-Alamos Archives -physics/0406126]
\bibitem {EY6}
R. Englman, A. Yahalom, M. Baer and A.M. Mebel "Some Experimental. and Calculational Consequences of Phases in Molecules with Multiple Conical Intersections" International Journal of Quantum Chemistry, 92, 135-151 (2003).
\bibitem {EY7}
R. Englman \& A. Yahalom, "Phase Evolution in a Multi-Component System", Physical Review A, 67, 5, 054103-054106 (2003). [Los-Alamos Archives -quant-ph/0406195]
\bibitem {EY8}
R. Englman \& A. Yahalom, "Generalized "Quasi-classical" Ground State of an Interacting Doublet" Physical Review B, 69, 22, 224302 (2004). [Los-Alamos Archives - cond-mat/0406725]
\bibitem {Spflu}
A. Yahalom "The Fluid Dynamics of Spin".  Molecu\-lar Physics, Published online: 13 Apr 2018.\\
 http://dx.doi.org/10.1080/00268976.2018.1457808\\ (arXiv:1802.09331 [physics.flu-dyn]).
\bibitem{Clebsch1}
Clebsch, A., Uber eine allgemeine Transformation der hydrodynamischen Gleichungen.
{\itshape J.~reine angew.~Math.}~1857, {\bf 54}, 293--312.
\bibitem{Clebsch2}
Clebsch, A., Uber die Integration der hydrodynamischen Gleichungen.
{\itshape J.~reine angew.~Math.}~1859, {\bf 56}, 1--10.
\bibitem {Davidov}
B. Davydov\index{Davydov B.}, "Variational principle and canonical
equations for an ideal fluid," {\it Doklady Akad. Nauk},  vol. 69,
165-168, 1949. (in Russian)
\bibitem {Eckart}
C. Eckart\index{Eckart C.}, "Variation\index{variational
principle} Principles of Hydrodynamics \index{hydrodynamics},"
{\it Phys. Fluids}, vol. 3, 421, 1960.
\bibitem {Bertherton}
F.P. Bretherton "A note on Hamilton's principle for perfect fluids," Journal of Fluid Mechanics / Volume 44 / Issue 01 / October 1970, pp 19 31 DOI: 10.1017/S0022112070001660, Published online: 29 March 2006.
\bibitem{Herivel}
J. W. Herivel  Proc. Camb. Phil. Soc., {\bf 51}, 344 (1955)
\bibitem{Serrin}
J. Serrin, {\it \lq Mathematical Principles of Classical Fluid
Mechanics'} in {\it Handbuch der Physik}, {\bf 8}, 148 (1959)
\bibitem{Lin}
C. C. Lin , {\it \lq Liquid Helium'} in {\it Proc. Int. School Phys. XXI}
(Academic Press)  (1963)
\bibitem{Seliger}
R. L. Seliger  \& G. B. Whitham, {\it Proc. Roy. Soc. London},
A{\bf 305}, 1 (1968)
\bibitem{LynanKatz}
D. Lynden-Bell and J. Katz "Isocirculational Flows and their Lagrangian and Energy principles",
Proceedings of the Royal Society of London. Series A, Mathematical and Physical Sciences, Vol. 378,
No. 1773, 179-205 (Oct. 8, 1981).
\bibitem{KatzLyndeb}
J. Katz \& D. Lynden-Bell 1982,{\it Proc. R. Soc. Lond.} {\bf A 381} 263-274.
\bibitem{YahLyndeb}
Asher Yahalom and Donald Lynden-Bell "Variational Principles for Topological Barotropic Fluid Dynamics" ["Simplified Variational Principles for Barotropic Fluid Dynamics" Los-Alamos Archives - physics/ 0603162] Geophysical \& Astrophysical Fluid Dynamics. 11/2014; 108(6). DOI: 10.1080/03091929.2014.952725.
\bibitem{LyndenB}
Lynden-Bell, D., Consequences of one spring researching with Chandrasekhar. {\itshape Current Science} 1996, {\bf 70}(9), 789--799.
\bibitem{Lamb H.}
Lamb, H., {\itshape Hydrodynamics}, 1945 (New York: Dover Publications).
\bibitem{Moff2}
H.K. Moffatt, \index{Moffatt H.K.} "The degree of knottedness of tangled vortex lines," {\it J. Fluid Mech.}, vol. 35, 117, 1969.
\bibitem{Schrodinger} E. Schr\"{o}dinger, Ann. d. Phys. {\bf 81} 109 (1926).
 English translation appears in E. Schr\"{o}dinger, {\it Collected Papers in Wave
 Mechanics} (Blackie and Sons, London, 1928) p. 102
  \bibitem{Fisher}
 R. A. Fisher {\it Phil. Trans. R. Soc. London} {\bf 222}, 309.
 \bibitem {Mandel}
L. Mandel and E. Wolf, {\it Optical Coherence and Quantum Optics} (University
Press, Cambridge, 1995) section 3.1
\bibitem {Fisherspin}
A. Yahalom "The Fluid Dynamics of Spin - a Fisher Information Perspective" arXiv:1802.09331v2 [cond-mat.] 6 Jul 2018. Proceedings of the Seventeenth Israeli - Russian Bi-National Workshop 2018 "The optimization of composition, structure and properties of metals, oxides, composites, nano and amorphous materials".
\bibitem {Frieden}
B. R. Frieden {\it Science from Fisher Information: A Unification} (Cambridge University Press, Cambridge, 2004)
\bibitem {Fisherspin2}
Asher Yahalom "The Fluid Dynamics of Spin - a Fisher Information Perspective and Comoving Scalars" Chaotic Modeling and Simulation (CMSIM) 1: 17-30, 2020.
\end {thebibliography}
\end {document}